\documentclass[prl,twocolumn,superscriptaddress]{revtex4}
\usepackage{amsmath}
\usepackage[dvips]{graphicx}
\usepackage{float}

\begin{document}

%\title{Comment on ``Excitation Chains at the Glass Transition'' -[Phys. Rev. Lett. \textbf{97}, 115704 (2006)]}
\title{Comparison of some recent Excitation Chain Arguments with the Random First Order Transition Theory of Supercooled Liquids and Experiment}
\date{\today}
\author{Jacob D. Stevenson}
\affiliation{Department of Physics and Department of Chemistry and Biochemistry,
University of California, San Diego, La Jolla, CA 92093}
\author{Peter G. Wolynes}
\affiliation{Department of Physics and Department of Chemistry and Biochemistry,
University of California, San Diego, La Jolla, CA 92093}
\affiliation{e-mail: pwolynes@ucsd.edu}

\begin{abstract}
We compare a recent excitation chain argument for the glass transition with the earlier random first order transition theory.  The key equation determining the activation barriers and size of cooperatively rearranging regions has the same scaling form in both approaches.  The random first order transition theory unambiguously predicts the coefficients in the equation giving results that agree with experiment vis a vis the correlation of activation barriers with thermodynamics and that also agree with experimental determination of correlation lengths following the prescription of Berthier et al.  The excitation chain approach, while containing more adjustable parameters accommodates those experimental findings only by using unphysical values for those adjustable parameters.
\end{abstract}

\maketitle

The nature of the cooperatively rearranging regions (CRRs) in supercooled molecular liquids and glasses has been difficult to establish owing to their extraordinarily slow dynamics and their mesoscopic size - larger than a few molecules but far from the scale easily probed by traditional scattering methods.  Langer has recently argued that the rearranging regions are excitation chains whose activation free energy can be estimated much like the droplet theory of the gas-liquid transition\cite{langer.2006}.  The latter imperfect analogy is shared by the random first order transition (RFOT) theory\cite{kirkpatrick.1987a, kirkpatrick.1989, xia.2000, lubchenko.2006} which has its roots in an analogy to spin glasses with discontinuous replica symmetry breaking\cite{mezard.2001} at the mean field level.  Although Langer seems to take issue with other theories based on random first order transitions, the key equation in both his approach and the RFOT theory relates the activation process to a free energy profile as a function of the size of the reconfigured region.  In both approaches the profile takes the same form.

\begin{equation}
F(N)=-a(T)N + b(T)N^{1/2}
\label{eq.1}
\end{equation}

At this level, the two approaches only differ in how the coefficients $a$ and $b$ are found.  In the RFOT theory, above the glass transition, $a(T)$ directly reflects the entropy driven nature of the activated transitions $a_{RFOT}(T)=TS_{c}(T)$, while $b(T)$ is given by a density functional calculation that reflects the entropy cost of immobilizing a particle.  Near the glass transition the $b(T)$ given by the RFOT density functional calculation has the explicit value 
$b_{RFOT}(T_K)=\frac{3\sqrt{3\pi}}{2}k_BT_K \log \frac{\alpha_L}{\pi e}$ 
where $\alpha_L \approx 100$ is the inverse square of the Lindemann ratio of the vibrational amplitude to the particle diameter and $T_K$ is the Kauzmann temperature where the configurational entropy is extrapolated to vanish.  Though obviously simplified, the density functional calculations suggest that $b_{RFOT}/T_K$ is a nearly universal quantity.  In Langer's excitation chain (EC) argument $a(T)$ depends on the shape entropy of a string-like excitation and an energy scale, but it is finally given by the expression $a_{EC}(T)=k_B\nu(T-T_0)$.  The coefficient $\nu$ is geometrical, representing the number of ways a link on the chain can be placed.  In the excitation chain approach $a(T)$ turns out to be proportional to the configurational entropy term
$TS_c(T)$, but is not equal to it.  $a(T)$ also does vanish at the temperature $T_0=T_K$.  Although no explicit microscopic calculation of $b(T)$ was provided, in the excitation chain argument this energy is accounted for by introducing an additional material dependent parameter of order unity, $\gamma_0$, which Langer describes as an ``inverse localization length associated with the fact that the excitation chains exist in the highly disordered environment characteristic of a molecular glass.''  Explicitly, the EC approach writes 
$b_{EC}(T) =  4(\pi/6)^{3/4}(k_B T_{int})^{1/4}(\gamma_0 k_B T_0^2 / T)^{3/4}$.

Just as the physical mechanisms leading to the linear $N$ term are different in the two approaches, the origin of the $N^{1/2}$ terms seems different in the two.  In the RFOT theory the $N^{1/2}$ scaling is due to a wetting effect much like what happens in a random field magnet\cite{villain.1985}.  In the EC approach the term is said to come from the self-interaction of excitation chains treated via a Flory\cite{flory.1953} entropy argument.

Near $T_g$, both the RFOT theory and Langer's analysis lead to the Vogel-Fulcher law for the activation free energy $\Delta F^{\ddagger}=DT_K/(T-T_K)$.  In addition, the RFOT calculation, because it contains no adjustable parameters, quantitatively predicts a relation between the coefficient in that law and the heat capacity change at the glass transition. to be precise $D=32k_B/\Delta C_P$.  The predicted relation for the activation enthalpy at $T_g$, $m$, agrees very well with experimental data for numerous substances\cite{stevenson.2005}.  The average slope of the data differs from the predicted one by only 10\% with an $R^2$ fit of 0.92.  The excitation chain formula can also be reconciled with this experimental finding if we are allowed to take the product of two parameters in the EC argument, $\gamma_0$ and $s_0$, which according to Langer is the configurational entropy of an unfrozen molecule multiplied by a domain wall thickness in molecular units, to be universal, i.e. $\gamma_0 s_0$ = constant.  

Since the key equation of the EC argument has the same N dependence as does RFOT theory above the glass transition, a practical distinction between the two versions of the scaling form can only be made on the basis of the coefficients in the expression (\ref{eq.1}).  Ultimately these coefficients determine, not only the barrier, but also the size of the rearranging regions.  Reconciling size and time scale, is, however, a key difficulty with any theory based on strings treated as unrenormalized elementary excitations, as pointed out by Lubchenko and Wolynes\cite{lubchenko.2004}.  They have argued that it is difficult for an elementary defect based theory to achieve very slow relaxations while still being consistent with the measured entropy density at the glass transition which is about $0.8k_B$ per particle.

This difficulty is confirmed when we contrast the quantitative comparison of the specific EC approach taken by Langer and the RFOT theory with experimental data.  The RFOT theory predicts (again without using adjustable parameters) that the correlation length at $T_g$, $\xi$, is a nearly universal quantity when measured in the size of the elementary movable units of the liquid (called ``beads'').  In these units $(\xi /a)_{RFOT}$ was predicted by Xia and Wolynes\cite{xia.2000} to be 4.5, a quantity independent of the fragility of the liquid.  

The correlation length at $T_g$  found from the EC calculation was given by Langer as
\begin{equation}
\xi_{EC} = \frac{3\lambda_g}{\pi \gamma_0}
\end{equation}
$\lambda_g$ is the dimensionless activation free energy corresponding to the glass relaxation time, $\lambda_g \approx 32$.  This expression thus depends on $\gamma_0$ alone, not the product $\gamma_0 s_0$.  In principle, therefore, strictly speaking, no universal form for $\xi$ should be predicted by the excitation chain argument.  Nevertheless, again, if we follow Langer and also assume that $\gamma_0$ is a universal number of order unity we would find $\xi_{EC} = 30/\gamma_0$, a quantity that as Langer states doesn't explicitly depend on fragility.  

There are a few fairly direct measurements of $\xi$ using nonlinear NMR\cite{sillescu.1999, richert.2002}, deviations from the Stokes-Einstein relation\cite{ediger.2000}, and single molecule imaging approaches\cite{russell.2000}.  These give values consistent with the RFOT prediction.  Because the size of rearranging regions determines the range of relaxation times, the size is also indirectly related to the stretching exponent in the stretched exponential form of the relaxation function.  Here again RFOT theory makes predictions that quantitatively agree with experiment\cite{xia.2001}.  On this issue, a related, more elegant analysis of the correlation lengths by Berthier et al.\cite{berthier.2005} is cited by Langer.  Berthier et al. show the dynamic correlation length defined in the 4 point function is bounded by a dynamic/ energetic correlation length which can be inferred from the relaxation function, $\phi(t)$ and the temperature dependence of the relaxation time.

\begin{figure}[tbp]
\includegraphics[width=0.48\textwidth]{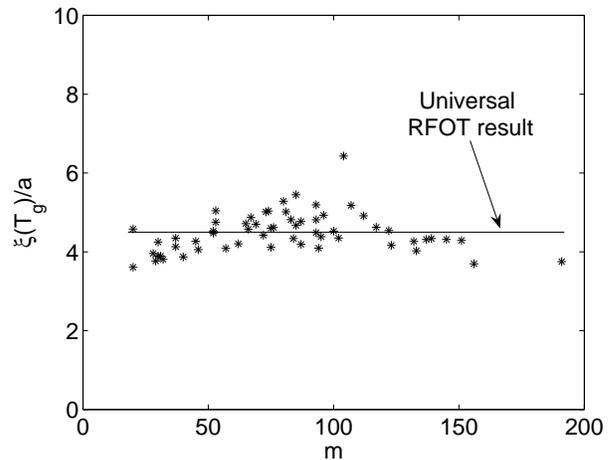}
\caption{\textbf{Dynamical correlation lengths} determined from the Berthier et al. relation\cite{berthier.2005} $\xi(T_g) /a=\left( \frac{\log(10)^2}{\pi e^2}\beta^2 m^2 k_B/\Delta C_P\right)^{1/3}$.  The values of $\Delta C_P$ were found using the relation $m=20.7 \Delta C_P/k_B$, where $\Delta C_P$ is given per independently rearranging molecular unit, or ``bead.''\cite{stevenson.2005}  All data were taken from B\"{o}hmer et al.\cite{bohmer.1993}}
\label{fig.1}
\end{figure}

This dynamic/energetic correlation length itself should correspond with the $\xi$ provided by RFOT theory or by the EC theory in that it deals with the thermodynamic/kinetic correlation that enters those analyses.  The values of $\xi$ already provided in the Berthier paper\cite{berthier.2005} cluster tightly around the RFOT prediction.  They are, however, considerably smaller than the value predicted by the EC calculation using the natural input value $\gamma_0=1$.  We have carried out the same analysis as Berthier et al. for 60 substances using other published data\cite{bohmer.1993}.  These results are shown in figure \ref{fig.1}, plotted against liquid fragility.  Again, the $\xi$ values cluster around the RFOT prediction and show rough universality independent of fragility.  The mean of $\xi$ is $4.47$ with a standard deviation $\pm 0.25$.  The results show very limited scatter, questioning the need for the additional adjustable parameters provided by the EC approach.  In addition, as we have seen, the naive estimate of $\gamma_0$ provided by Langer ($\gamma_0 = 1$), gives much larger correlation lengths than those inferred from experiment in this way, or from the direct measurements cited earlier. To fit both experimental correlations simultaneously, the free parameters in the EC theory, $\gamma_0$ and $s_0$, would both have to be adjusted so that the shared free energy profile (equation \ref{eq.1}) would essentially coincide with the RFOT prediction.  The value of $\gamma_0$ needed would then be 6.  This corresponds to a length associated with the disorder of 1/6 for the string.  It is hard to see how to reconcile this fractional value with Langer's physical picture of a chain of molecules in a disordered medium.  The elementary excitation chain argument as provided by Langer, unlike the earlier RFOT theory, seems inconsistent with experiment concerning the crucial question of the size of the cooperative regions.

Our criticism of the excitation chain approach is aimed only at the specific excitation chain theory put forward in reference \cite{langer.2006}, not at all string based arguments, which have their place in glass physics.  We have argued elsewhere that near the so-called crossover temperature where deviations from the Vogel-Fulcher law occur, cooperatively rearranging regions, in fact, do become fractal and could be rather roughly approximated as strings\cite{stevenson.2006} as seen in computer simulations\cite{donati.1998}.  Indeed that string theoretic calculation predicts an unexpected correlation between the crossover temperature signaling a deviation from Vogel-Fulcher behavior and fragility which is reasonably well confirmed experimentally\cite{stevenson.2005}.  A key feature of the RFOT based argument is that the cooperatively rearranging regions are predicted to be most string-like near the crossover and to become more compact at lower temperature.  Langer, contrarily, suggests that his argument implies the strings are shortest near the crossover and merely lengthen at low temperature, rather than changing shape.  This distinction concerning the shapes of cooperatively rearranging regions may allow experimental distinctions to be made using sufficiently refined imaging techniques in the laboratory.   We also note that string-like dislocations have long featured in some structural theories of glasses\cite{nelson.1983}.  Nussinov has shown that such models are not inconsistent with an RFOT theory treatment\cite{nussinov.2004}.  At the defect densities needed for real liquids these elementary dislocations would, of necessity, be highly renormalized, however.

\begin{acknowledgments}
We thank L. Berthier and G. Biroli for helpful, clarifying correspondence concerning reference \cite{berthier.2005}.  This work was supported by NSF Grant CHE0317017.
\end{acknowledgments}

\bibliography{/home/jake/latex_files/jakes_biblio.bib}

\end{document}